\documentclass[prl,twocolumn,showpacs,showkeys,amsfonts,
amssymb,amsmath,letterpaper]{revtex4}
\usepackage{times}
\usepackage{graphicx}
\usepackage[bookmarks=false]{hyperref}
\hypersetup{pdftitle={Separation of plastic deformations in polymers based 
on elements of general 
nonlinear theory},
pdfauthor={Sergei F. Lyuksyutov, Ruslan A. Sharipov}}
\begin{document}
\def\ty{\tilde y}
\title{Separation of plastic deformations in polymers\\ based on elements 
of general nonlinear theory}
\author{Sergei F. Lyuksyutov}
\affiliation{\href{http://nebula.physics.uakron.edu/dept/faculty/sfl.htm}
Departments of Physics and Polymer Engineering, The University of Akron, Akron OH 44325, USA}
\author{Ruslan A. Sharipov}
\affiliation{\href{http://www.bashedu.ru}
Department of Mathematics, Bashkir State University, 32 Frunze street, Ufa 450074, Russia}
\date{February 18, 2004}
\begin{abstract}We report a method for describing plasticity in a broad class of amorphous materials. The method is based on nonlinear (geometric) deformation theory allowing the separation of the plastic deformation from the general deformation tensor. This separation allows an adequate pattern of thermodynamical phenomena for plastic deformations in polymers to be constructed. A parameter, $\theta$ describing the stress relaxation rate of the material is introduced within the frame of this approach. Additionally, several experimental configurations to measure this parameter are discussed.   
\end{abstract}
\maketitle
Plastic and elastic deformation plays a critical role in the description of physical 
phenomena in Earthis crustal deformations \cite{EarthDef}, transfer of slurry, powdery 
and granular materials through pipelines \cite{Granular}, and, especially, in polymer dynamics \cite{PGdeGenes}. Although the theory of elastic deformation including elastic equilibrium, 
deflection and torsion of rods, bending of plates and shells is a classical and well 
developed field, a completed theory applied to plastic media is yet to be developed.
\par

The theories describing microscopic plastic deformations in metals is based on a born-death 
dislocation mechanism producing a shift of crystallographic grid. This have been developed by 
E.~Orowan, M.~Polanyi, G.~I.~Taylor, J.~M.~Burgers, F.~C.~Frank, W.~T.~Read, R.~E.~Pierels,
P.~B.~Hirsch, W.~C.~Dash, Yu.~A.~Osipyan et al (\cite{PierelsOsipyan}). In solids, the shear transformation zone theory (STZ)  
has been developed by Falk and Langer \cite{FalkLanger,LobovskyLanger}. However, a complete 
theory of plastic deformation in polymer materials has not developed so far. Physical 
behavior of polymers at the nanoscale is important from the fundamental point of view to understand deformation processes, evolution of homogeneous or heterogeneous 
nanostructures, and the thermal history of monomer architecture 
\cite{Importance1,Importance2}.
Understanding the deformation behavior in polymers would 
eventiually produce novel nanostructure formation techniques based on nano-deformations.
\par

Polymers are probably most suitable materials in which plastic deformation can be studied experimentally. Recently, Lyuksyutov and Vaia with co-authors reported nanopattering technique based on localized Joule heating of a thin polymer films 
\cite{NatureMaterials2, ApplPhysLett83}. A biased atomic force microscope (AFM) tip produces
an electric current flow through the polymer film resulting in localized Joule heating of 
the polymer above its glass transition temperature due to electronic breakdown through the film. Polarization and electrostatic attraction 
of softened polymer toward the AFM tip in the presence of a strong ($10^9$-$10^{10}$ Vm$^{-1}$) non-uniform electric 
field produce raised or depressed nanostructures (10-50 nm width, and 0.1-100 nm height) in a broad 
class of polymers of different physical-chemical properties. This technique, named AFM-based electrostatic nanolithography (AFMEN), can be applied to the study of plastic deformations since both a liquid and a plastic 
solid polymer phases coexist during the process. The breakdown during AFMEN is a critical factor causing film softening, and polymer mass transfer as a result.
Recent experimental data indeed produce the evidence of nanostructure formation in polymer  that cannot be explained by the electronic 
breakdown. The most likely reason for 
polymer nanostructure formation when an AFM tip of 20-50 nm in 
diameter moves above the surface is the plastic deformation of polymer molecules through triboelectrification mechanism. No electronic breakdown is required to deform the polymer surface in this case. An exact analytical solution, based on the method of images, has been obtained for the description of the electric field between an atomic force microscope (AFM) tip and a thin dielectric polymer film (30 nm thick) spin-coated on a conductive substrate. Three different tip shapes are found to produce electrostatic pressure above the plasticity threshold in the polymers up to 50 MPa \cite{PRB}.
Should such a technique, based on plastic deformations in polymer materials, be 
developed further, it would create an alternative to existing nanopattering techniques effective tool, for patterning on nanoscale. 
There has been a lot of activity in in last few years in developing these techniques including: 1) direct resist lithography developed by Schaeffer et al \cite{Schaeffer} based on the competition of Van der Waal's and Laplace forces on polymer-air interface in strong electric field; and 2) hierarchic nanostructure formation based on electrodynamic instability in bilayer polymer films developed by Russell et al \cite{Russell}. The only industrial prototype for nanostructure formation in polymers 
called MILLIPEDE developed by Vettinger et al \cite{IBMResDevelop} is based on thermal-mechanical lithography developed by Mamin and Rugar yearlier in the 90s. 
\cite{ThermoWriting}. The authors of the MILLIPEDE project predict a replacement of ferromagnetic memories with based on polymers in the next 20 years. 
\par
Experimental verification of the existence of plastic deformation on nanoscale in AFM-tip-polymer-metal system \cite{PRB} requires a theoretical model to describe them explicitly.
An important question is how to describe a polymer surface undergoing 
deformations near the glass transition point when two phases exist. There are two approaches 
for describing polymers in the liquid-solid phase deformed under external forces. The first, would be based on intensive mathematical description of this process through solution of the Navier-Stokes equation for a steady flow of non-Newtonian 
incompressible liquid with non-slip boundary conditions \cite{bookTaner}. Although this 
approach may be ultimately correct, it lacks generality for the description of polymer deformation 
at the nanoscale. The second approach would be based on time-dependent geometry of 
elastic and plastic deformations using the elements of differential geometry 
and tensor analysis. In this paper we consider this approach.
\par
The goal of this letter is the separation of plastic deformations 
within a general nonlinear deformation tensor.  A general 
method for the separation would be to build into a framework of balance equations 
traditionally used in the description of dynamics and thermodynamics of moving 
continuous media. The model we present, would be useful for the description 
of solid-liquid substances including organic molecules like DNA and long polymer molecules. 
\par
\begin{figure}[!hbp]
\includegraphics[]{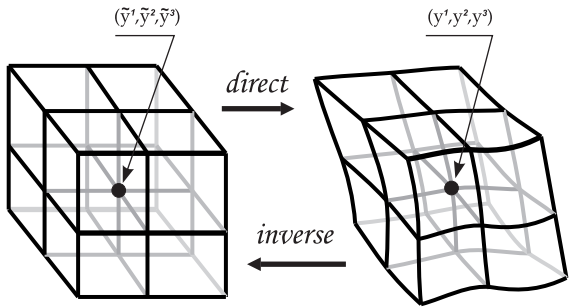}
\caption[Figure 1]{\ Schematic presentation of the direct and inverse transformations on the deformation map. A point is transformed}
\label{FigMaps}
\end{figure}

The technique is based on the tensor calculus associated with curvilinear 
coordinates system. The basic concept for moving continuous media description 
is presented through the deformation map. The map transforms the point $(\ty^1\!,\,\ty^2\!,\,\ty^3)$ of the non-deformed 
medium to its current actual position $(y^1\!,\,y^2\!,\,y^3)$. Direct and inverse 
deformation maps can be presented through the following sets of functions:
\begin{equation}
\label{EqMaps}
\hskip -2em
\left\{\begin{aligned}
&y^1=y^1(\ty^1,\ty^2,\ty^3),\\
&y^2=y^2(\ty^1,\ty^2,\ty^3),\\
&y^3=y^3(\ty^1,\ty^2,\ty^3),
\end{aligned}
\right.
\quad
\left\{\begin{aligned}
&\ty^1=\ty^1(y^1,y^2,y^3),\\
&\ty^2=\ty^2(y^1,y^2,y^3),\\
&\ty^3=\ty^3(y^1,y^2,y^3),
\end{aligned}
\right.
\end{equation}
The direct and inverse transformations of the deformation map are presented in Figure 1 ~\ref{FigMaps}.
The partial derivatives of the mapping functions \eqref{EqMaps} define two Jacobi matrices 
$\tilde S^i_j=\partial y^i/\partial\ty^j$ and $\tilde T^i_j=\partial\ty^i/\partial\ty^j$.
The non-linear deformation tensor can be defined using one of them as pesented below:
\begin{equation}
\label{DefTensor}
\hskip -2em
G_{ij}=\sum^3_{r=1}\sum^3_{s=1}g_{rs}(\ty^1,\ty^2,\ty^3)\,\tilde T^r_i\,\tilde T^s_j,
\end{equation}
where $g_{rs}$ is the metric tensor arising through the useage of curvilinear 
coordinates \cite{SflShr,MattHeil}. In Cartesian coordinates 
$g_{rs}$ is presented by the unit matrix; $\tilde T^r_i$, and $\tilde T^s_j$ are two Jacobi marices.
\par 
The tensor \eqref{DefTensor} is a quantitative measure of the deformation 
at given point of curvilinear coordinate system. Should a continuous medium be represented as a 
collection of infinetisemal cubes, then the tensor
$G_{ij}$ completely describes how the edges and angles of these
cubes contract, elongate and distort
(as presented in Figure~\ref{FigMaps}). Differentiating the formula \eqref{DefTensor}we arrive at the following evolution differential equation for the tensor $G_{ij}$:
\begin{equation}
\label{DefTensorEq}
\hskip -2em
\begin{aligned}
\frac{\partial G_{ij}}{\partial t}&+\sum^3_{r=1}v^r\,\nabla_{\!r}G_{ij}
=\\
&-\sum^3_{r=1}\nabla_{\!i}v^r\,G_{rj}-\sum^3_{r=1}G_{ir}\,\nabla_{\!j}v^r.
\end{aligned}
\end{equation}

In order to describe plastic media, we separate the general deformation 
tensor into two tensors: $\hat G$, and $\check G$ are two parts of plastic deformation tensor of;  $\hat G_{kq}$ is the tensor of elastic deformation: 

\begin{equation} 
\label{SeparFormula}
\hskip -2em
G_{ij}=\sum^3_{k=1}\sum^3_{q=1}\check G^{\,k}_i\,\hat G_{kq}
\,\check G^{\,q}_j.
\end{equation}

The choice of representing the plastic deformation tensor in two parts (instead of one as for elastic tensor) is related to the symmetry of the general tensor. 
The separation is the major step for the following consideration.
The plastic deformation occurs in plastic materials that include pitch-like dense sticky
liquids, solid amorphous materials, e.\,g.
glasses, and polymers. These materials resist to the plastic deformation like
solids and can flow like liquids. Unlike metals, where plastic 
deformation produces dislocations resulting in disorder of the
crystalline grid, in our consideration, the plastic deformation does not 
change the material stucture.
\par

The separation produces separate evolution equations for its plastic and elastic parts. The evolution equation of the plastic deformation $\check G$ can be presented as:

\begin{equation} 
\label{PlasticDefEq}
\hskip -2em
\begin{aligned}
\frac{\partial\check G^{\,k}_i}{\partial t}&+
\sum^3_{r=1}v^r\,\nabla_{\!r}\check G^{\,k}_i=
\sum^3_{r=1}\check G^{\,r}_i\,\nabla_{\!r}v^k\\
&-\sum^3_{r=1}\nabla_{\!i}v^r\,\check G^{\,k}_r-\sum^3_{r=1}
\theta^{\,k}_r\,\check G^{\,r}_i.
\end{aligned}
\end{equation} 
Here, as in equation \eqref{DefTensorEq} $v^r$ is the velocity of medium; $\nabla$ is the spatial derivative; $\theta$ is the physical parameter of continuous medium associated with the relaxation rate of elastic deformation into plastic. 

Although the tensor $\theta^{\,k}_r$ is not symmetric, it can be obtained from the symmetric tensor $\theta_{ij}$ through the standard procedure of index raising. Using equations
\eqref{DefTensorEq}, \eqref{SeparFormula}, \eqref{PlasticDefEq} we derive the evolution equation of elastic deformation. The evolution of elastic deformation is an important factor in determining the stress of a medium.

\begin{equation} 
\label{ElasticDefEq}
\hskip -2em
\begin{gathered}
\frac{\partial\hat G_{kq}}{\partial t}+\sum^3_{r=1}v^r\,
\nabla_{\!r}\hat G_{kq}=-\sum^3_{r=1}\nabla_{\!k}v^r\,
\hat G_{rq}\\
-\sum^3_{r=1}\hat G_{kr}\,\nabla_{\!q}v^r
+\sum^3_{r=1}
\theta^{\,r}_k\,\hat G_{rq}+\sum^3_{r=1}\hat G_{kr}
\,\theta^{\,r}_q.
\end{gathered}
\end{equation} 
\par 

Equations \eqref{PlasticDefEq} 
and \eqref{ElasticDefEq} must be complemented by three balance equations (three conservation 
laws for the flows of mass, momentum , and energy). The first is the mass balance equation:
\begin{equation}
\label{MassBalance}
\hskip -2em
\frac{\partial\rho}{\partial t}
+\sum^3_{k=1}\nabla_{\!k}(\,\rho\,v^k)=0.
\end{equation}

where $\rho$ is the density; $v^k$ is the velocity tensor.
The next two equations are the momentum, and the energy balance 
equations:
\begin{align}
\label{MomentumBalance}
\hskip -2em
\frac{\partial(\,\rho\,v^i)}{\partial t}
+\sum^3_{k=1}\nabla_{\!k}\Pi^{ik}=f^i,\\
\label{EnergyBalance}
\hskip -2em
\frac{\partial}{\partial t}\!\left(\frac{\rho\,|\mathbf v|^2}{2}+
\rho\,\varepsilon\right)+\sum^3_{k=1}\nabla_{\!k}w^k=e.
\end{align}

Here $f^i$ is the density of external forces acting on the medium, $e$ is the power of the forces; $v^1$, $v^2$, $v^3$ are
the components of velocity vector $\mathbf v$; $\varepsilon$ is the specific
inner thermal energy; $\Pi^{ik}$ and $w^k$ are the momentum, and energy flows determined by the following formulae:

\begin{align*}
&\Pi^{ik}=\rho\,v^i\,v^k-\sigma^{ik}-\sum^3_{j=1}\sum^3_{q=1}\eta^{ikjq}\,v_{jq},\\
&\begin{aligned}
w^k&=\frac{\rho\,|\mathbf v|^2}{2}\,v^k+\rho\,\varepsilon\,v^k
-\sum^3_{i=1}v_i\,\sigma^{ik}\\
&-\sum^3_{i=1}\sum^3_{j=1}\sum^3_{q=1}v_i\,\eta^{ikjq}\,v_{jq}
-\sum^3_{i=1}\nabla_{\!i}T\,\varkappa^{ik},
\end{aligned}
\end{align*}

where $\sigma^{ik}$ is the stress tensor; 
$\eta^{ikjq}$ is the viscosity tensor; $\varkappa^{ik}$ is the heat conductivity
tensor and $v_{jq}$ is the tensor of velocity gradients:
\begin{equation*}
v_{ij}=\frac{\nabla_{\!i}v_j+\nabla_{\!j}v_i}{2}.
\end{equation*}
\par

The parameter $\theta$ has the units of inverse time: $(sec^{-1})$, and  
determines the stress relaxation rate. The external forces produce stress and elastic deformation as a responce to this stress. In solely elastic material the responce is constant for constant deformation, and the parameter $\theta$ is zero. In plastic materials $\theta$ is non-zero and the exponential reduction of elastic response is observed. For constant 
general deformation $G$, the elastic component $\hat G$ in \eqref{SeparFormula}
decreases, while the plastic part $\check G$ increases. 
This scenario is presented in Figure 2.

\begin{figure}[!h]
\includegraphics[]{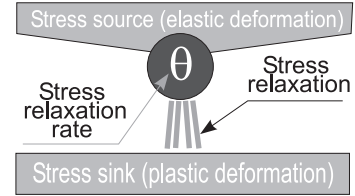}
\caption[Figure 2]{\ Conceptual presentation of parameter $\theta$. Elastic deformation produces stress relaxation resulting in plastic deformation through the conservation
laws.}
\label{FigSink}
\end{figure}
\par
\noindent

Further analysis of this model indicates that for materials in which the plastic 
deformation does not change the structure and produces no stress, it can excluded from 
consideration. This can be formulated through the "`forgetting 
principle"'. Suppose the medium evolves from an initial 
state at time $t=0$, to the
intermediate state, at $t=t_0$, and then continues evolution 
for $t>t_0$.
The principle asserts that if the intermediate 
state total deformation of medium was solely plastic, then the
medium evolution occurs without memory of the intermediate state.

\par
Equations \eqref{PlasticDefEq}-\eqref{EnergyBalance}
form a complete set describing the evolution of plastic medium
in the frames of this model. They are used to describe the thermodynamics
of plastic deformations. The specific internal thermal energy of the material depends on the entropy and on the elastic deformation. The differential of the energy can be written as:

\begin{equation}
\label{DifEpsilon}
\hskip -2em
d\varepsilon=T\,ds-\sum^3_{i=1}\sum^3_{j=1}\frac{\bar\sigma^{ij}\,d\hat G_{ij}}
{2\,\rho}
\end{equation}

Here, $\bar\sigma^{ij}$ is an auxiliary tensor related to the stress tensor 
through the following formula:

\begin{equation}
\label{AuxiliarySigma}
\hskip -2em
\sigma^{ij}=\sum^3_{r=1}\hat G^{\,i}_r\,\bar\sigma^{rj}.
\end{equation}

Using equations \eqref{DifEpsilon}, and \eqref{AuxiliarySigma} the time 
derivative of $\varepsilon$ and its gradient can be found. Substituting them
into equation \eqref{EnergyBalance} and using the equations 
\eqref{PlasticDefEq}-\eqref{MomentumBalance}, we obtain the equation for specific entropy \cite{SflShr}:
 
\begin{equation}
\label{EntropyProd}
\hskip -2em
\begin{gathered}
\frac{\partial(\,\rho\,s)}{\partial t}+\sum^3_{k=1}\nabla_{\!k}\!
\left(\rho\,s\,v^k-\smash{\sum^3_{i=1}}\frac{\nabla_{\!i}T\,
\varkappa^{ik}}{T}\right)\\
=\sum^3_{i=1}\sum^3_{j=1}\frac{\sigma^{ij}
\,\theta_{ij}}{T}
+\sum^3_{i=1}\sum^3_{k=1}\frac{\nabla_{\!i}T\,\varkappa^{ik}\,
\nabla_{\!k}T}{T^2}\\
+\sum^3_{i=1}\sum^3_{k=1}\sum^3_{j=1}\sum^3_{q=1}
\frac{v_{ik}\,\eta^{ikjq}\,v_{jq}}{T}
\end{gathered}
\end{equation}

The three terms in the right hand side of equation \eqref{EntropyProd} 
describe three different mechanisms of entropy production: 
plasticity, heat transfer, and viscosity. 
Second and third terms of \eqref{EntropyProd} are 
typical for the visco-elastic solids. The term 
\begin{equation}
\label{PlasticityTerm}
\sum^3_{i=1}\sum^3_{j=1}\frac{\sigma^{ij}
\,\theta_{ij}}{T}
\end{equation}
responsible for plasticity is new one. It is non-zero indicating 
the growth of entropy on the way to equilibrium. Elastic deformation produces stress leading to plastic deformation, which in turn, results in additional entropy production as shown in \eqref{EntropyProd}.

\par
Once introduced, the material parameter $\theta$ should 
be measured experimentally. The parameter can be found using calorimetric measuremens, as can be seen from the temperature dependence of the term \eqref{PlasticityTerm}. Alternatively, this parameter can be measured through the static bending of a polymer, or dynamic flow of the material. These two possible approaches for $\theta$ measurements are presented in Figure 3 below.
\begin{figure}[!h]
\includegraphics[]{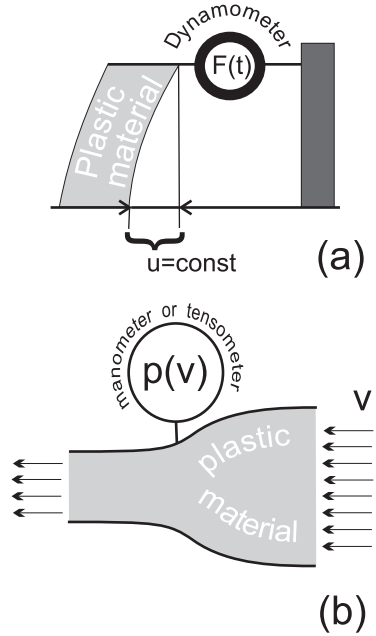}
\caption[Figure 3]{\ (a)Static configuration of $\theta$ mesurement. A polymer sample of rectangular shape is subjected to shear bending through the application of lateral force $\mathbf F(t)$. The parameter $\theta$ can be determined by keeping $u=\operatorname{const}$ and measuring the force $\mathbf F$ as a function of time; (b)Dynamic configuration is comprised of the polymer pumped through the inverse nozzle. The pressure $p(v)$ on the walls of nozzle is measured as a function of velocity of incoming plastic flow, from which the components of tensor $\theta$ can be calculated. }
\label{FigStatic}
\end{figure}
\par
In summary, we have suggested an approach for the separation of plastic deformation within general nonlinear tensor deformation. This separation allows for a robust description of the material's plasticity using the relaxation rate introduced through the parameter  $\theta$.  This approach can be also used to describe experimental data of polymer deformation on nanoscale.

The authors acknowledge the support from National Research Council grant through the COBASE program. The authors wish to thank R. R. Mallik for the useful comments.

\end{document}